\documentclass[
 aip,
 amsmath,amssymb,
 preprint,%
]{revtex4-1}

\usepackage[utf8]{inputenc}
\usepackage[T1]{fontenc}
\usepackage{mathptmx}
\usepackage{graphicx}
\usepackage{dcolumn}
\usepackage{bm}
\usepackage{booktabs}
\usepackage{float}
\usepackage[caption=false]{subfig}
\usepackage[dvipsnames]{xcolor}
\definecolor{BLUE}{rgb}{0,0,1}
\usepackage{soul}
\usepackage[colorlinks=true,linkcolor=cyan,urlcolor=blue,citecolor=blue,anchorcolor=black]{hyperref}
\makeatletter
\newcommand{\smallsym}[2]{#1{\mathpalette\make@small@sym{#2}}}
\newcommand{\make@small@sym}[2]{%
  \vcenter{\hbox{$\m@th\downgrade@style#1#2$}}%
}
\newcommand{\downgrade@style}[1]{%
  \ifx#1\displaystyle\scriptstyle\else
    \ifx#1\textstyle\scriptstyle\else
      \scriptscriptstyle
  \fi\fi
}
\makeatother

\def\selectlanguage#1{}

\begin{document}

\title{Hybrid kinetic-MHD simulation on the formation of runaway electron current plateau during a current quench process}

\author{Weimin Fu}
\affiliation{School of Physics, Huazhong University of Science and Technology, Wuhan, Hubei, 430074, China}
\affiliation{State Key Laboratory of Advanced Electromagnetic Technology, International Joint Research Laboratory of Magnetic Confinement Fusion and Plasma Physics, School of Electrical and Electronic Engineering, Huazhong University of Science and Technology, Wuhan, Hubei, 430074, China}

\author{Ping Zhu}
\email{zhup@hust.edu.cn}
\affiliation{State Key Laboratory of Advanced Electromagnetic Technology, International Joint Research Laboratory of Magnetic Confinement Fusion and Plasma Physics, School of Electrical and Electronic Engineering, Huazhong University of Science and Technology, Wuhan, Hubei, 430074, China}
\affiliation{Department of Nuclear Engineering and Engineering Physics, University of Wisconsin-Madison, Madison, Wisconsin, 53706, United States of America}

\date{\today}

\begin{abstract}
Runaway electron (RE) generation is of great concern for high-current tokamak operations. A full-f particle-in-cell (PIC) model for RE dynamics has been developed and coupled with the 3D nonlinear extended magnetohydrodynamic (MHD) model implemented in the NIMROD code. Our model accounts for RE generation using analytical source terms and advances the RE motion along guiding-center (GC) orbits. The model was employed to simulate the formation of RE current plateau during a disruption, in which the plasma current becomes dominated by the RE current. The model agrees well with several RE codes based on fluid models when the RE GC drifts are ignored. For highly relativistic REs, the grad-$B$ and curvature drifts can play a significant role in the RE generation and motion due to the increase in the safety factor profile during the current quench process.
\end{abstract}


\maketitle

\section{Introduction}
Disruptions are sudden losses of thermal and magnetic energy confinement that can terminate tokamak discharges and impose extreme heat and electromagnetic loads on plasma-facing components (PFCs)~\cite{Ratynskaia_2025}. A typical disruption process consists of a sequence of phases: thermal quench (TQ) and current quench (CQ), which in some cases includes the formation of a runaway-electron (RE) current plateau. During the TQ, the plasma thermal energy is rapidly lost. The subsequent CQ is characterized by the decay of the bulk plasma current as the plasma cools down and the corresponding resistivity increases. As a consequence, a large toroidal electric field is induced, which can accelerate a small population of electrons to relativistic energies and trigger RE generation. In high-current tokamaks, the REs generated during CQ phase may carry a substantial fraction of the pre-disruption plasma current and form a long-lived RE current plateau. If such an RE beam is lost in an uncontrolled manner, its highly localized energy deposition can cause severe damage or even deep melting of PFCs, resulting in unacceptably long machine downtimes. Therefore, understanding the formation, evolution, and transport of RE current plateaus is essential for disruption mitigation and RE control in present and future tokamak devices.

Experiments on DIII-D~\cite{Paz-Soldan_2019}, JET~\cite{reux_runaway_2015}, ASDEX Upgrade~\cite{Gobbin_2024}, and J-TEXT~\cite{CHEN_2022,Wang_2026} have applied several methods to the mitigation of RE generation and damage, including the massive impurity injection and the utilization of runaway electron mitigation coils (REMCs). In the meantime, numerical models have been developed to interpret these experiments and to predict RE behavior in future devices. The 3D fully nonlinear MHD codes NIMROD~\cite{sainterme_resistive_2024}, M3D-$C^1$~\cite{liu_self-consistent_2021,zhao_simulation_2024}, and JOREK~\cite{bandaru_simulating_2019} have implemented fluid RE models that include the parallel advection and the Dreicer and avalanche generation mechanisms~\cite{dreicer_2electron_1959,rosenbluth_theory_1997}. However, fluid descriptions often simplify or ignore the kinetic orbit effects of REs, such as those from grad-$B$ and curvature drifts, whereas existing particle models do not always include self-consistent RE source terms associated with Dreicer and avalanche generation~\cite{lopez_2025,Bergström_2025}.

In this paper, we present a full-$f$ RE particle-in-cell (PIC) model that accounts for both RE generation mechanisms and guiding-center drift effects. The analytical source terms for the Dreicer and avalanche mechanisms are implemented either by adjusting marker weights or by increasing the number of simulation markers. The RE guiding-center motion is advanced including grad-$B$ and curvature drifts. The ensemble average of RE motions is coupled to the evolution of MHD fields in NIMROD. The model is benchmarked against existing fluid RE models in NIMROD, JOREK, and DREAM. We further use the model to study the self-consistent formation of RE current plateaus during current quench processes and to examine how RE drift effects influence the RE distribution, current evolution, and possible loss of high-energy REs.

The rest of this paper is organized as follows. In Section~\ref{sec:model}, we introduce the PIC RE model, including the source terms, and its implementation in the NIMROD code. In Section~\ref{sec:benchmark}, we benchmark the model against the NIMROD, JOREK, and DREAM~\cite{hoppe_dream_2021} fluid RE models. Section~\ref{sec:drift} describes the RE drift effects found in disruption simulations. Finally, in Section~\ref{sec:dis} we provide the discussion and conclusion.

\section{The PIC RE model for NIMROD code}
\label{sec:model}
Our model is based on the PIC model for energetic-particles (EPs)~\cite{kim_hybrid_2004} implemented in the NIMROD code{\cite{sovinec_nonlinear_2004}}. In this PIC model, a marker's physical coordinates $(R, Z)$ in the cylindrical coordinate system and its logical coordinates $(p, q)$ in a 2D finite element in the poloidal plane are connected by
\begin{equation}
R=\sum_{i=1}^{m^2} R_i\, N_i(p, q), \qquad
Z=\sum_{i=1}^{m^2} Z_i\, N_i(p, q). 
\end{equation}
Here, $N_i(p,q)$ denotes the 2D finite element basis functions, $(R_i,Z_i)$ are the physical space coordinates of the finite element nodes, and $m$ is the number of nodes along each side of the finite element. Markers are advanced using the physical fields evaluated at their locations. In the full-$f$ method, the relation between the distribution of markers $f_m$ and the distribution of real particles $f_p$ can be written as~\cite{lu_full_2023}
\begin{equation}
f{_p}(\boldsymbol{x},t)
\approx C_{\mathrm{m2p}} \sum_{{j}=1} ^{{N_m}} w_{{j},\mathrm{tot}} (t)\,
 {\frac{\delta\!\left[\boldsymbol{x} - \boldsymbol{x}_{{j}}(t)\right]}{\mathcal{J}}},
\end{equation}
where $\boldsymbol{x}$ represents the phase-space coordinates, $\delta$ is the Dirac delta function, $\mathcal{J}$ is the corresponding Jacobian, {$C_{\mathrm{m2p}} \equiv N_p/N_m$, and $N_m$ and $N_p$} denote the total numbers of markers and real particles, respectively. For each marker, in the absence of explicit source operations, the total weight carried by marker $j$ is conserved as
\begin{equation}
 {w_{{j},\mathrm{tot}}} (t)
=\frac{1}{C_{\mathrm{m2p}}}\, {\frac{f_p(\boldsymbol{x},t)}{f_m(\boldsymbol{x},t)}} = {\mathrm{const}}.
\label{eq:3}
\end{equation}

 The RE number density and parallel current density are defined as velocity-space moments of the RE distribution function ($f_\mathrm{RE}\equiv f_p$ for RE particles)
\begin{equation}
 n_{\mathrm{RE}}(\boldsymbol{R},t)
=
\int f_{\mathrm{RE}}(\boldsymbol{R},\boldsymbol{v},t)\,\mathrm{d}^{3}v,
\qquad
J_{\mathrm{RE},\parallel}(\boldsymbol{R},t)
=
-e\int v_{\parallel}f_{\mathrm{RE}}(\boldsymbol{R},\boldsymbol{v},t)\,\mathrm{d}^{3}v.
\end{equation}
Since NIMROD employs finite elements in the poloidal plane and a finite Fourier series in the toroidal direction, these moments are projected onto the finite-element--Fourier basis. Substituting the full-$f$ marker representation into the above moment definitions and carrying out the phase-space integration over the Dirac delta functions gives the nodal coefficients for RE number and parallel current density:
 \begin{equation}
 \begin{aligned}
\left\{n_{\mathrm{RE}},J_{\mathrm{RE},\parallel}\right\}^{i_R,i_Z,n}
&\equiv
\int_{\Omega}\mathrm{d}V\,
\left\{n_{\mathrm{RE}}(\boldsymbol{R},t),J_{\mathrm{RE},\parallel}(\boldsymbol{R},t)\right\}
N_{i_R}(R)N_{i_Z}(Z)e^{-i n\phi}
\\
&=
C_{\mathrm{m2p}}
\sum_{{j}=1}^{N_m}
w_{{j},\mathrm{tot}}
\left\{1,-e v_{\parallel,{j}}\right\}
N_{i_R}(R_{{j}})N_{i_Z}(Z_{{j}})e^{-i n\phi_{{j}}}.
\end{aligned}
\label{eq:density and current}
\end{equation}
Here, {$n$ denotes the toroidal Fourier mode number,} $\boldsymbol{R}$ is the {real-space} coordinate, $N_{i_\ast}$ denote the 1D finite element basis functions, and $(R_{{j}},Z_{{j}},\phi_{{j}})$ and {$v_{\parallel,{j}}$} denote the position and {parallel} velocity associated with marker $j$, respectively. Since NIMROD uses a finite Fourier series {for discretization} in the toroidal direction, we adopt the particle-in-Fourier method in Eq.~(\ref{eq:density and current})~\cite{mitchell_efficient_2019}. These drift terms are identical to those implemented in the test particle model in NIMROD~\cite{izzo_runaway_2011}. The GC orbits of particles are described by

\begin{equation}
    \begin{aligned}
&\mathrm{d} R=\frac{v_{\|} B_R}{B} \mathrm{~d} t+\frac{1}{B^2}[\vec{E} \times \vec{B}]_R \mathrm{~d} t,\\
&\begin{aligned}
\mathrm{d} Z & =\frac{v_{\|} B_Z}{B} \mathrm{~d} t+\frac{1}{R} \frac{\gamma m_{e} v_{\perp}^2}{2 e B} \mathrm{~d} t+\frac{1}{R} \frac{\gamma m_{e} v_{\|}^2}{e B} \mathrm{~d} t \\
& +\frac{1}{B^2}[\vec{E} \times \vec{B}]_Z \mathrm{~d} t,
\end{aligned}\\
&\mathrm{d} \phi=\frac{v_{\|} B_\phi}{R B} \mathrm{~d} t+\frac{1}{R B^2}[\vec{E} \times \vec{B}]_\phi \mathrm{d} t,
\end{aligned}
\label{eq:drift eq}
\end{equation}
where $v_\parallel$ is the parallel velocity, $v_\bot$ is the perpendicular velocity of the RE, $m_e$ is the electron rest mass, {$\gamma=1/\sqrt{1-(v/c)^2}$} is the Lorentz factor, and a large aspect ratio approximation is used to avoid calculating gradients of the magnetic field. The original PIC model in NIMROD included a full orbit Boris algorithm solver and a gyro-kinetic predictor-corrector solver. We applied the Cash-Karp fifth-order Runge-Kutta method \cite{cash_1990} to push particles in the same magnetic field.  We couple the {PIC-RE model} to the three-dimensional nonlinear MHD code NIMROD through a modified Ohm's law and the momentum equation:

\begin{gather}
	\frac{\partial \rho}{\partial t}
	+\nabla \cdot(\rho \boldsymbol{V})=0, 
	\\
	\rho\left(
	\frac{\partial \boldsymbol{V}}{\partial t}
	+\boldsymbol{V} \cdot \nabla \boldsymbol{V}
	\right)
	=
	\boldsymbol{J}\times\boldsymbol{B}
	-\nabla p
	-\boldsymbol{J}_{\mathrm{RE},\perp}\times\boldsymbol{B},
	\label{eq:mhd_momentum}
	\\
	\frac{1}{\Gamma-1}
	\left(
	\frac{\partial p}{\partial t}
	+\boldsymbol{V} \cdot \nabla p
	\right)
	=
	-p \nabla \cdot \boldsymbol{V}
	-\nabla\cdot\boldsymbol{q},
	\label{eq:pressure}
	\\
	\boldsymbol{q}
	=
	-\kappa_\parallel
	\boldsymbol{b}\boldsymbol{b}\cdot\nabla T
	-\kappa_\perp
	\left(
	\boldsymbol{I}
	-
	\boldsymbol{b}\boldsymbol{b}
	\right)
	\cdot\nabla T,
	\label{eq:heat_flux}
	\\
	\frac{\partial \boldsymbol{B}}{\partial t}
	=
	-\nabla \times \boldsymbol{E},
	\\
	\boldsymbol{J}
	=
	\frac{1}{\mu_0} \nabla \times \boldsymbol{B},
	\\
	\boldsymbol{E}
	=
	-\boldsymbol{V}\times\boldsymbol{B}
	+
	\eta\left(
	\boldsymbol{J}
	-
	\boldsymbol{J}_{\mathrm{RE}}
	\right).
	\label{eq:ohm}
\end{gather}
where $\boldsymbol{J}_{\mathrm{RE}}$ denotes the runaway-electron current,
$\boldsymbol{b}=\boldsymbol{B}/B$, $\boldsymbol{I}$ is the identity tensor,
$T$ is the plasma temperature, and $\kappa_\parallel$ and
$\kappa_\perp$ are the parallel and perpendicular thermal conductivities,
respectively. The resistivity associated with the runaway current is assumed
to be zero, and Ohmic heating is neglected. 

The total runaway current is decomposed into the parallel and perpendicular components,
\begin{equation}
    \boldsymbol{J}_{\mathrm{RE}}
    =
    \boldsymbol{J}_{\mathrm{RE},\parallel}
    +
    \boldsymbol{J}_{\mathrm{RE},\perp},
\end{equation}
where
\begin{equation}
    \boldsymbol{J}_{{\mathrm{RE}},\parallel}
    =
    J_{{\mathrm{RE}},\parallel}\boldsymbol{b},
    \qquad
    \boldsymbol{b}=\frac{\boldsymbol{B}}{B}.
\end{equation}
Since $\boldsymbol{J}_{\mathrm{RE},\parallel}\times\boldsymbol{B}=0$, the parallel RE current does not contribute directly to the Lorentz force in the momentum equation. Therefore, only the perpendicular RE current appears in Eq.~(\ref{eq:mhd_momentum}).

 Following the guiding-center formulation of Ref.~\cite{liu_hybrid_2026}
, the drift and magnetization currents can be combined to express the perpendicular RE current in terms of the gyrotropic RE pressure. Terms proportional to the RE charge density, including the corresponding $\boldsymbol{E}\times\boldsymbol{B}$ contribution, are neglected because the RE number density is much smaller than the bulk-plasma density. The perpendicular RE current can be represented by the anisotropic RE pressure as
\begin{equation}
   \boldsymbol{J}_{\mathrm{RE},\perp}
   =
   \frac{1}{B}
   \left(
   p_{{\mathrm{RE}},\parallel}
   -
   p_{{\mathrm{RE}},\perp}
   \right)
   \nabla \times \boldsymbol{b}
   +
   \frac{1}{B}
   \boldsymbol{b}\times \nabla p_{{\mathrm{RE}},\perp}.
   \label{eq:jre_perp}
\end{equation}
 Here, $p_{\mathrm{RE},\parallel}$ and $p_{\mathrm{RE},\perp}$ denote the parallel and perpendicular components of the gyrotropic RE pressure, respectively. Substituting Eq.~(\ref{eq:jre_perp}) into the RE force term gives
\begin{equation}
   \boldsymbol{J}_{\mathrm{RE},\perp}\times\boldsymbol{B}
   =
   \left(
   p_{\mathrm{RE},\parallel}
   -
    p_{\mathrm{RE},\perp}
   \right)
   \left(\nabla \times \boldsymbol{b}\right)\times \boldsymbol{b}
   +
   \left(
   \boldsymbol{b}\times \nabla p_{\mathrm{RE},\perp}
   \right)\times \boldsymbol{b}.
   \label{eq:jre_cross_b}
\end{equation}
 Thus equivalently, the momentum equation can be written explicitly in terms of the RE pressure perturbations as
\begin{equation}
\begin{split}
    \rho\left(\frac{\partial \boldsymbol{V}}{\partial t}
    +\boldsymbol{V}\cdot\nabla\boldsymbol{V}\right)
    =
    \boldsymbol{J}\times\boldsymbol{B}
    -\nabla p
    &-
    \left(
    p_{\mathrm{RE},\parallel}
    -
    p_{\mathrm{RE},\perp}
    \right)
    \left(\nabla \times \boldsymbol{b}\right)\times \boldsymbol{b}
    \\
    &-
    \left(
    \boldsymbol{b}\times \nabla p_{\mathrm{RE},\perp}
    \right)\times \boldsymbol{b}.
\end{split}
\label{eq:mhd_momentum_pressure_coupling}
\end{equation}
Note that the RE contribution to the momentum equation is included through the pressure coupling, where the force associated with the perpendicular RE current is represented by the anisotropic RE pressure. The full RE current, including the parallel component, is retained in the modified Ohm's law, whereas only the perpendicular component contributes to the momentum balance.

One straightforward way to calculate the RE source is to use a Monte Carlo method to simulate the conversion of thermal electrons to REs. However, this method is quite expensive. In order to save computational cost, we use an analytical method to {account for the runaway electron source}. We adopt the Dreicer source through the Connor-Hastie equation~\cite{connor_relativistic_1975}:
\begin{equation}
    S_{D} = Cn{_e} \nu_{ee}\epsilon_d^{-\frac{3}{16}\left(1+Z_\mathrm{eff}\right)}e^{\left(-\frac{1}{4}\epsilon_d^{-1}-\left(1+Z_\mathrm{eff}\right)^{1/2}\epsilon_d^{-1/2}\right)} e^{\left[-\frac{T_e}{m_ec^2}\left(\frac{1}{8}\epsilon_d^{-2}+\frac{2}{3}\left(1+Z_\mathrm{eff}\right)^{1/2}\epsilon_d^{-3/2}\right)\right]}   
    \label{eq:C-H}
\end{equation}
Here $\nu_{ee}$ is the thermal electron-electron collision frequency{, $n_e$ is the number density of thermal electrons,} $C$ is a factor  usually set to unity, {$\epsilon_d=E_\parallel/E_D$} and 

\begin{equation}
E_{D}=\frac{n_{e} e^3 \ln \Lambda}{4 \pi \epsilon_0^2 T_{e}}
\end{equation}
 is the Dreicer field.
The avalanche term is adopted from the Rosenbluth-Putvinski model~\cite{rosenbluth_theory_1997}:

\begin{equation}
    S_A=n_{{\mathrm{RE}}}\nu_{{\mathrm{RE}}e}\frac{\epsilon_c - 1}{\ln \Lambda}\sqrt{\frac{\pi \varphi}{3\left(Z_\mathrm{eff}+5\right)}}\times\left(1-\epsilon_c+\frac{4\pi\left(Z_\mathrm{eff}+1\right)^2}{3\varphi\left(Z_\mathrm{eff}+5\right)\left(\epsilon^2+4/\varphi^2-1\right)}\right)^{-1/2}
    \label{eq:avalanche}
\end{equation}
 where $\nu_{\mathrm{RE}e}$ is the RE-electron collision frequency, $\varphi=(1+1.46 \sqrt{\epsilon}+1.72 \epsilon)^{-1}$ represents the collision effect averaged along the RE trajectory, with $\epsilon=r / R$ being the inverse aspect ratio, and  $\epsilon_c=E_{\|} / E_c$ with the critical electric field $E_c$ given by
\begin{equation}
E_c=\frac{n_e e^3 \ln \Lambda}{4 \pi \epsilon_0^2 m_e c^2}.
\end{equation}
In PIC models, usually the markers have a finite size in order to make the resulting self-consistent field smoother and to reduce numerical artifacts. 

Equation~(\ref{eq:density and current}) gives the
finite-element-Fourier projections for point markers. The pointwise
finite-element assignment in the poloidal plane can be written as
\begin{equation}
\begin{aligned}
N_{i_R}(R_j)N_{i_Z}(Z_j)
&=
\int_{\Omega_{RZ}}
R'\,\mathrm{d}R'\,\mathrm{d}Z'\,
\delta_{RZ}\!\left(R'-R_j,Z'-Z_j\right)
N_{i_R}(R')N_{i_Z}(Z').
\end{aligned}
\label{eq:point assignment}
\end{equation}
Here, the poloidal-plane delta function $\delta_{RZ}(R,Z)$ is normalized with respect to the cylindrical volume measure such that
\begin{equation}
\begin{aligned}
&\int_{\Omega_{RZ}}
R'\,\mathrm{d}R'\,\mathrm{d}Z'\,
\delta_{RZ}\!\left(R'-R_j,Z'-Z_j\right)
F(R',Z')
\\
&\qquad =
F(R_j,Z_j)
\end{aligned}
\label{eq:delta normalization}
\end{equation}
for an arbitrary smooth function $F$. Substituting Eq.~(\ref{eq:point assignment}) into Eq.~(\ref{eq:density and current}) gives the equivalent point-marker form
\begin{equation}
\begin{aligned}
\left\{
n_{\mathrm{RE}},
J_{\mathrm{RE},\parallel}
\right\}^{i_R,i_Z,n}
&=
C_{\mathrm{m2p}}
\sum_{j=1}^{N_m}
w_{j,\mathrm{tot}}
\left\{
1,-e v_{\parallel,j}
\right\}
e^{-in\phi_j}
\\
&\quad\times
\int_{\Omega_{RZ}}
R'\,\mathrm{d}R'\,\mathrm{d}Z'\,
\delta_{RZ}\!\left(R'-R_j,Z'-Z_j\right)
N_{i_R}(R')N_{i_Z}(Z').
\end{aligned}
\label{eq:point deposition integral}
\end{equation}

For a finite-size marker, the point-particle delta function in
Eq.~(\ref{eq:point deposition integral}) is replaced by the
two-dimensional particle shape function $S_l(R,Z)$, i.e.
\begin{equation}
\delta_{RZ}\!\left(R'-R_j,Z'-Z_j\right)
\longrightarrow
S_l\!\left(R'-R_j,Z'-Z_j\right),
\label{eq:delta to shape}
\end{equation}
where $l$ is a non-negative integer denoting the order of the particle
shape function in the poloidal plane. Under the standard B-spline
assignment convention, $l=0$, $1$, and $2$ correspond to the
nearest-grid-point (NGP), cloud-in-cell (CIC), and
triangular-shaped-cloud (TSC) assignments, respectively~\cite{birdsall_plasma_2004}. The particle
shape function is normalized with respect to the cylindrical volume
measure according to
\begin{equation}
\int_{\Omega_{RZ}}
R'\,\mathrm{d}R'\,\mathrm{d}Z'\,
S_l\!\left(R'-R_j,Z'-Z_j\right)=1.
\label{eq:shape normalization}
\end{equation}
Applying Eq.~(\ref{eq:delta to shape}) to
Eq.~(\ref{eq:point deposition integral}) gives the
finite-element--Fourier projections for finite-size markers:
\begin{equation}
\begin{aligned}
\left\{
n_{\mathrm{RE}},
J_{\mathrm{RE},\parallel}
\right\}_{l}^{i_R,i_Z,n}
&=
C_{\mathrm{m2p}}
\sum_{j=1}^{N_m}
w_{j,\mathrm{tot}}
\left\{
1,-e v_{\parallel,j}
\right\}
e^{-in\phi_j}
\\
&\quad\times
\int_{\Omega_{RZ}}
R'\,\mathrm{d}R'\,\mathrm{d}Z'\,
S_l\!\left(R'-R_j,Z'-Z_j\right)
N_{i_R}(R')N_{i_Z}(Z').
\end{aligned}
\label{eq:finite-size projection}
\end{equation}
For compactness, the integrated particle-to-finite-element assignment
is denoted as the weight
\begin{equation}
W_l^{i_R,i_Z}(j)
\equiv
\int_{\Omega_{RZ}}
R'\,\mathrm{d}R'\,\mathrm{d}Z'\,
S_l\!\left(R'-R_j,Z'-Z_j\right)
N_{i_R}(R')N_{i_Z}(Z'),
\label{eq:integrated assignment weight}
\end{equation}
so that Eq.~(\ref{eq:finite-size projection}) becomes
\begin{equation}
\begin{aligned}
\left\{
n_{\mathrm{RE}},
J_{\mathrm{RE},\parallel}
\right\}_{l}^{i_R,i_Z,n}
&=
C_{\mathrm{m2p}}
\sum_{j=1}^{N_m}
w_{j,\mathrm{tot}}
\left\{
1,-e v_{\parallel,j}
\right\}
W_l^{i_R,i_Z}(j)e^{-in\phi_j}.
\end{aligned}
\label{eq:shape function}
\end{equation}
In another word, $W_l^{i_R,i_Z}(j)$ is the integrated discrete assignment weight from
marker $j$ to the finite-element node $(i_R,i_Z)$. In
general, it is determined jointly by the particle shape function, the
finite-element basis functions, and the cylindrical volume measure.

The quantities in Eqs.~(\ref{eq:density and current}) and
(\ref{eq:shape function}) are finite-element-Fourier projections,
consistent with Eq.~(7) of Ref.~\cite{lu_full_2023}. When nodal
coefficients are required in the numerical implementation, the
finite-element mass matrix is mass lumped, giving
\begin{equation}
\left\{
n_{\mathrm{RE}},
J_{\mathrm{RE},\parallel}
\right\}_{\mathrm{nodal},l}^{i_R,i_Z,n}
=
\frac{1}{V_{i_R,i_Z}}
\left\{
n_{\mathrm{RE}},
J_{\mathrm{RE},\parallel}
\right\}_{l}^{i_R,i_Z,n},
\label{eq:mass lumped nodal moments}
\end{equation}
where
\begin{equation}
V_{i_R,i_Z}
\equiv
\sum_{k_R,k_Z}
M_{(i_R,i_Z),(k_R,k_Z)}
\label{eq:effective volume}
\end{equation}
is the lumped mass, or effective volume, associated with the
finite-element node $(i_R,i_Z)$, and
\begin{equation}
\begin{aligned}
M_{(i_R,i_Z),(k_R,k_Z)}
&\equiv
\int_{\Omega}
\mathrm{d}V\,
N_{i_R}(R)N_{i_Z}(Z)
N_{k_R}(R)N_{k_Z}(Z)
\end{aligned}
\label{eq:finite element mass matrix}
\end{equation}
is the finite-element mass matrix. 
If the finite-element basis satisfy the partition-of-unity
property,
\begin{equation}
\sum_{k_R,k_Z}
N_{k_R}(R)N_{k_Z}(Z)
=
1,
\label{eq:partition of unity}
\end{equation}
then the lumped mass can also be written as
\begin{equation}
\begin{aligned}
V_{i_R,i_Z}
&=
\sum_{k_R,k_Z}
M_{(i_R,i_Z),(k_R,k_Z)}
\\
&=
\int_{\Omega}
\mathrm{d}V\,
N_{i_R}(R)N_{i_Z}(Z)
\sum_{k_R,k_Z}
N_{k_R}(R)N_{k_Z}(Z)
\\
&=
\int_{\Omega}
\mathrm{d}V\,
N_{i_R}(R)N_{i_Z}(Z).
\end{aligned}
\label{eq:effective volume integral}
\end{equation}

 To model the source term in PIC models, usually we add more markers in cells to represent more real particles. However, RE generation is a long-time process, and the distribution range of runaway electron density is quite large, and the generated runaway current is  a very small portion of the plasma current, ranging from several mA to more than $10\, \mathrm{kA}$. This is quite complicated and computationally expensive if we add markers in every single step to achieve the effects of runaway electron sources, and the computer memory could also be a problem because there may be too many markers in the end. In order to implement the runaway electron source term while keeping the computational cost low, we adopt the NGP method, meaning that the markers' spatial shapes are $\delta$ functions and particles are deposited to the nearest grid point of {each cell. Such a method} has been used in gyrokinetic PIC simulations~\cite{PARKER2002520}. Thus we are able to model the runaway source  by increasing the weight $w_{j,\mathrm{tot}}$ for each marker. In this way, we can easily count the $\Delta w_{j,\mathrm{tot}}$ for each marker nearest a grid point. In practice, we can  generate REs by either adding markers into the domain, or just increasing the weight of markers in certain areas,
\begin{equation}
    \Delta w_{j\mathrm{,tot}}=\frac{VS_r\Delta t}{N}
    \label{eq:22}
\end{equation}
Here $S_r$ includes the Dreicer and avalanche source contributions. Thus, {whereas} Eq.~(\ref{eq:3}) describes the weight conservation during marker advection,  Eq.~(\ref{eq:22}) represents one possible way used to account for RE generation without increasing the total number of markers. Although the RE PIC module is implemented in the 3D nonlinear MHD framework of NIMROD, the simulations presented in this paper are restricted to axisymmetric  configurations unless otherwise specified. This allows the source implementation and drift effects to be examined without additional transport induced by 3D MHD perturbations.

\section{Disruption simulation benchmarks}
\label{sec:benchmark}
 The above PIC-RE model for runaway generation  is benchmarked with the fluid-RE models in NIMROD and JOREK RE codes. We assume the distribution of runaway electrons in velocity space as
 \begin{equation}
f_0=\delta{\left(\boldsymbol{v}-c\boldsymbol{b}\right)},
\end{equation}
and ignore the grad-$B$ and curvature drifts, consistent with the fluid-RE model. Here we run an artificial quench case, in which the core temperature drops due to large perpendicular thermal conductivity, which mimics a thermal quench. The equilibrium code CHEASE~\cite{lutjensCHEASECodeToroidal1996a} is used to generate the equilibrium which is close to the case in Ref. \cite{bandaru_simulating_2019} where the initial total current $I_p=0.71\, \mathrm{MA}$, the initial central temperature $T_0=1.7$ keV, and the number density $n_0=1 \times 10^{20}\, \mathrm{m}^{-3}$. The benchmark simulations are performed on $32\times32$ bicubic finite elements in poloidal plane with only the axisymmetric $n=0$ toroidal Fourier component. We also set the effective RE parallel speed as $10^{-3} c$.

\begin{figure*}[!htbp]
    \centering{
        \includegraphics[width=0.8\textwidth]{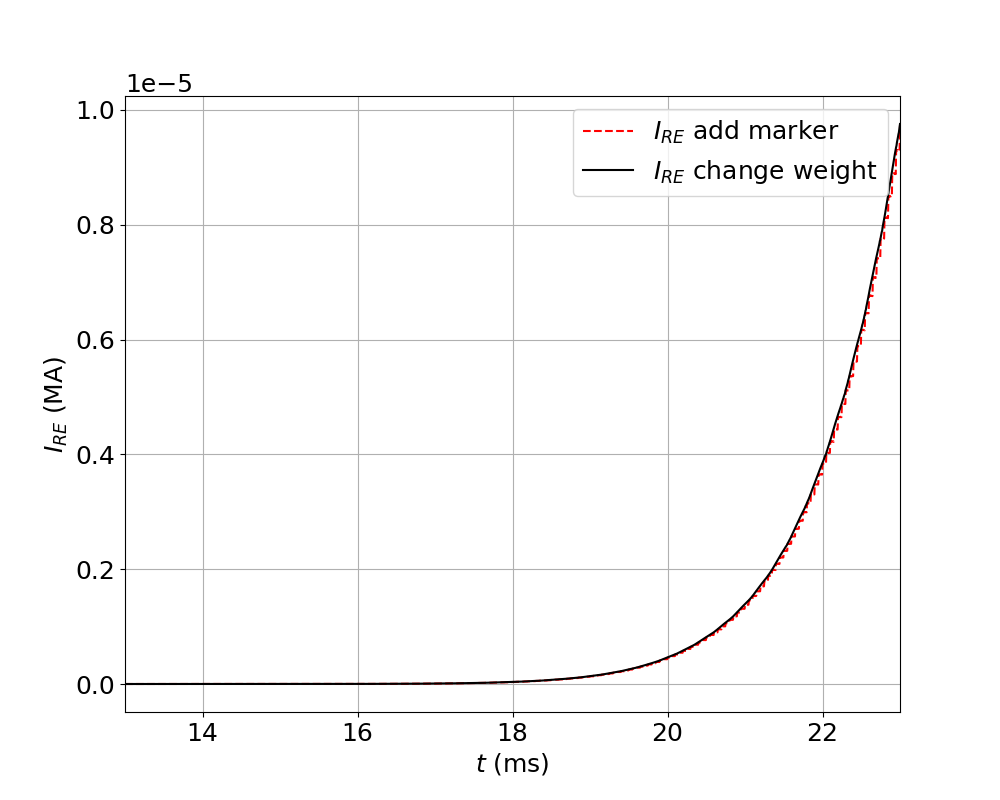}
    }
    \caption{The total runaway current {evolutions as functions of time} with {sources computed} using the {"}add marker{"} and {"}change weight{" methods}.}
        \label{fig: addmarker vs changeweight}
\end{figure*}

The methods of changing weight and adding markers for the RE generation are tested and compared using $1\times10^5$ markers  by changing their weight versus adding approximately $1024$ markers to cells every 10 steps respectively. The result shown in Figure \ref{fig: addmarker vs changeweight} indicates that these two methods are consistent when neglecting the influence of the RE distribution in momentum space, but the weight-changing method is much faster than the marker-adding method when the simulation period is long. In order to save computational resource, the following simulations are conducted using the weight-changing method.

\begin{figure*}[!htbp]
    \centering
    \subfloat[Time evolution comparison]{
        \includegraphics[width=0.8\textwidth]{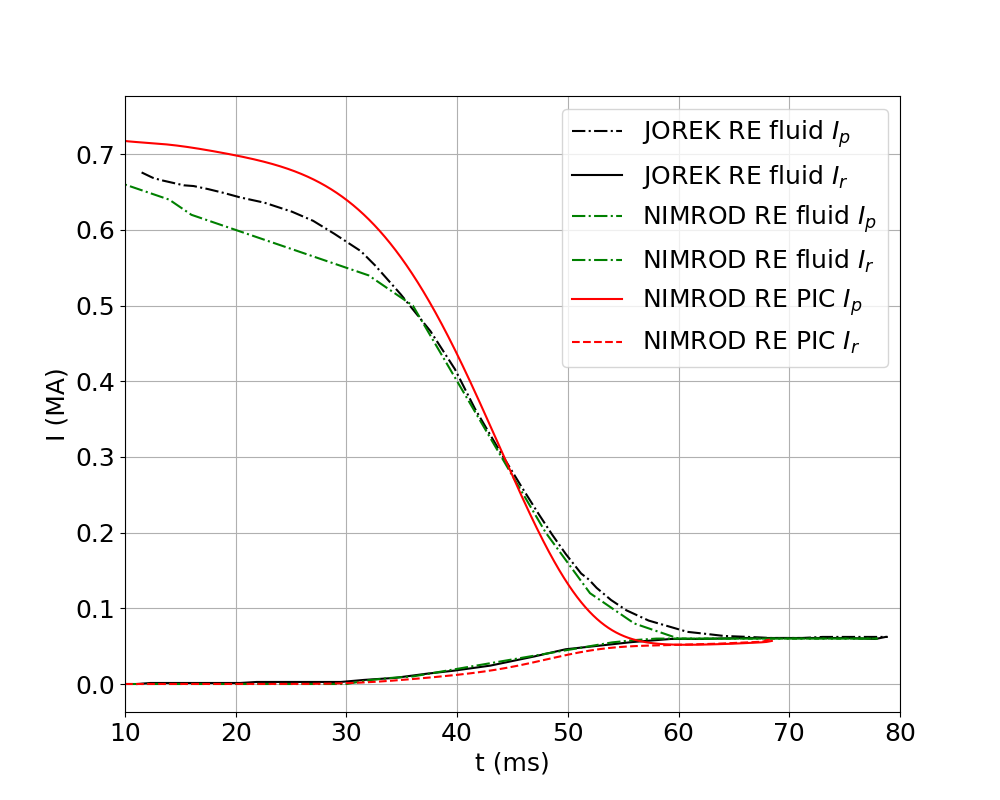}
        \label{fig: total current NIMROD JOREK}
    }
    \hfill
    \subfloat[Current profile comparison]{
        \includegraphics[width=0.8\textwidth]{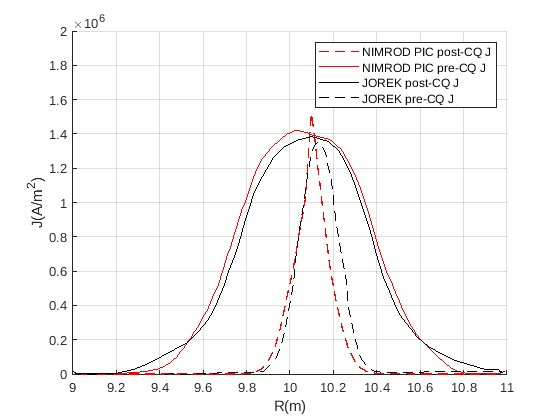}
        \label{fig: JOREK current profile}
    }
    \caption{(a) Comparison of the total current {evolutions as functions of time, and} (b) {comparison} of the {pre-CQ} and {post-CQ} current profiles {as functions of major radius computed using various codes}.}
    \label{fig:JOREK}
\end{figure*}

\begin{figure*}[!htbp]
    \centering
    \subfloat[Time evolution comparison]{
        \includegraphics[width=0.8\textwidth]{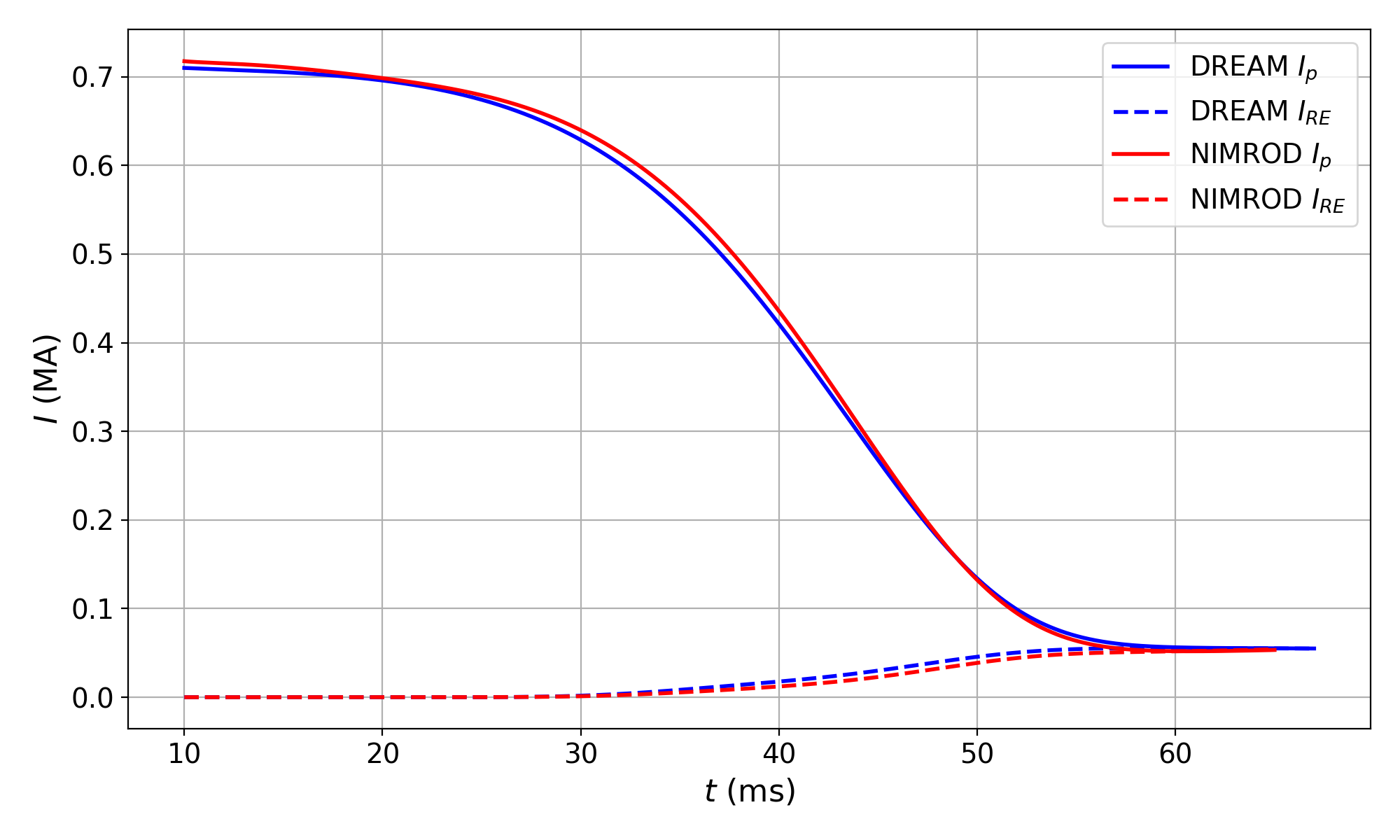}
        \label{fig: total current NIMROD DREAM}
    }
    \hfill
    \subfloat[Current profile comparison]{
        \includegraphics[width=0.8\textwidth]{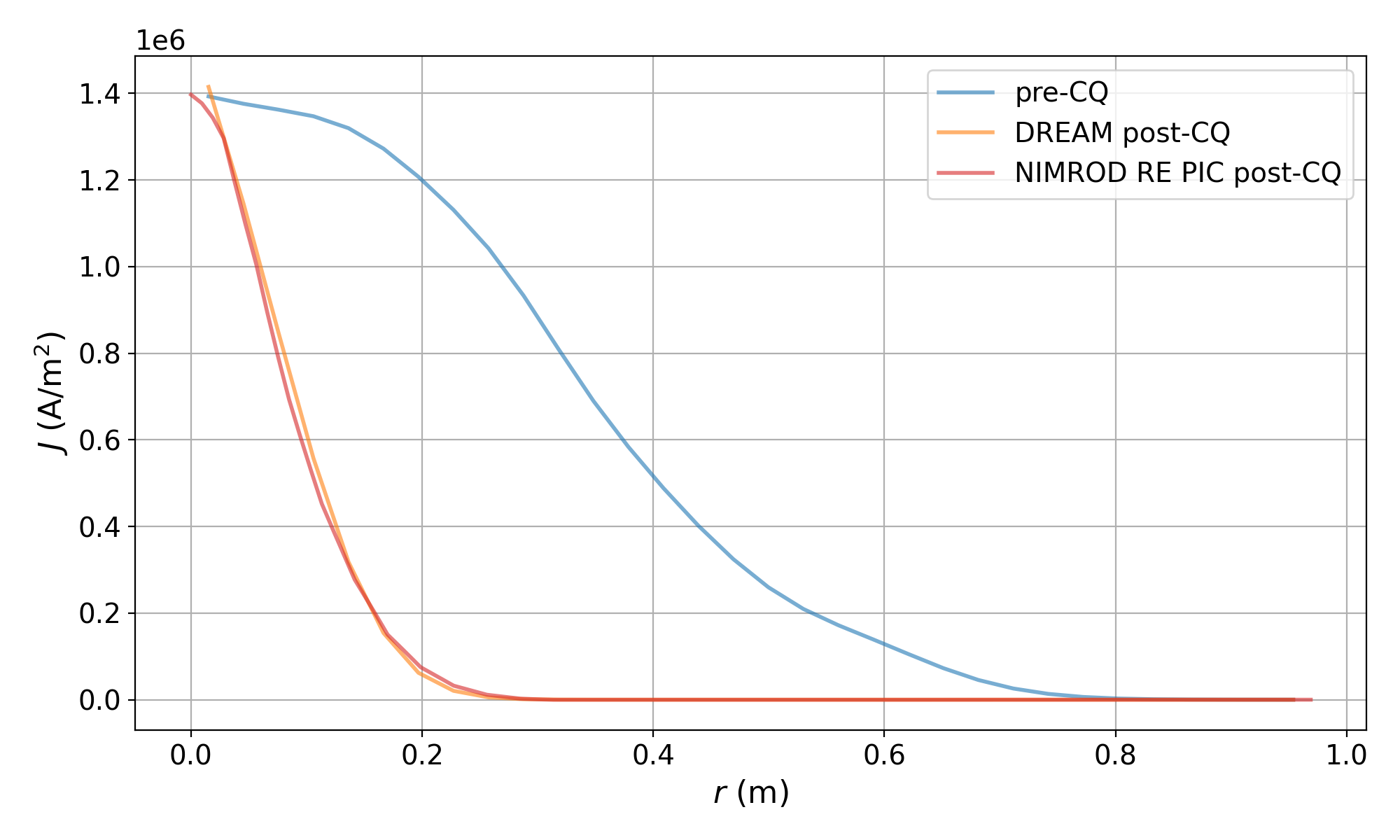}
        \label{fig: DREAM current profile}
    }
    \caption{(a) Comparison of total {plasma and RE} current {evolutions} {as functions of time and} (b) {comparison} of pre- and post-disruption current profiles {as functions of minor radius computed using various codes}.}
    \label{fig:DREAM}
\end{figure*}

Figure \ref{fig: total current NIMROD JOREK} and Figure \ref{fig: JOREK current profile} show good agreement among the NIMROD fluid\cite{Sainterme2022} and PIC-RE models and the JOREK fluid-RE model~\cite{bandaru_simulating_2019}. Our model is also benchmarked with the one-dimensional RE code DREAM, similar to the way Bandaru et al.~\cite{bandaru_simulating_2019} have benchmarked the JOREK RE model with the one-dimensional RE code GO~\cite{smith_runaway_2006}. We directly input our equilibrium into DREAM using its tool EQDSK to provide a similar initial condition. We run the same case with the prescribed temperature profile from NIMROD using the fluid model {in DREAM. By running} the DREAM cases in \verb|AVALANCHE_MODE_FLUID| and \verb|DREICER_RATE_CONNOR_HASTIE_NOCORR| modes,  equations similar to Equation (\ref{eq:C-H}) and Equation (\ref{eq:avalanche}) {are used} to calculate the runaway electron generation rate. Figure \ref{fig: total current NIMROD DREAM} and Figure \ref{fig: DREAM current profile} show a good {agreement between the NIMROD PIC-RE} model and the DREAM fluid model.

\section{Effects of RE drifts in disruption simulations}
\label{sec:drift}
The runaway electron's mass {$m=m_e\gamma$} can be quite large {at relativistic speed.}
 According to Equation (\ref{eq:drift eq}), the drift effects could be quite significant for these relativistic electrons. Here we  assume that the REs' momentum space distribution is
\begin{equation}
    f=\delta\left(\boldsymbol{p}-\boldsymbol{p}_0\right),
\end{equation}
and their pitch is set as $v_\bot/v_\parallel=0.15$.

\begin{figure*}[!htbp]
    \centering{
        \includegraphics[width=1.0\textwidth]{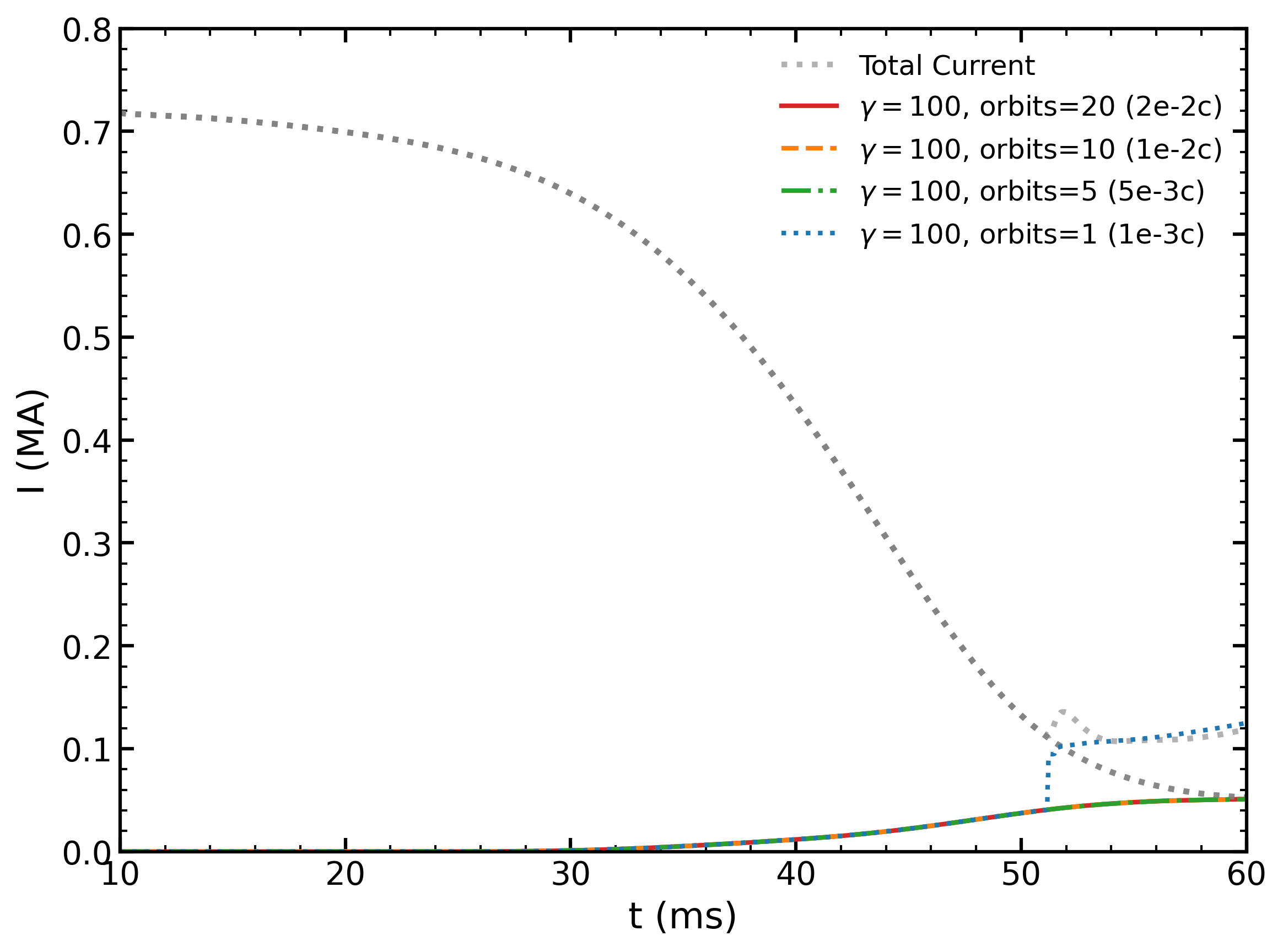}
    }
    \caption{The total plasma and RE current evolutions as functions of time for various initial RE convection speed and orbit averaging. The abrupt jump in the $1\times 10^{-3}c$ case is caused by insufficient orbit averaging and is therefore regarded as a numerical artifact. Converged behavior is obtained when at least five orbits, corresponding to $5\times 10^{-3}c$, are used. }
        \label{fig: total_comparison}
\end{figure*}

Before examining the drift effects associated with different RE energies, we first performed a numerical sensitivity test on the effective RE convection speed and orbit averaging used in the current-deposition scheme. Orbit averaging is applied to reduce the statistical noise associated with particle-to-mesh deposition. During one MHD time step, each RE marker is advanced through multiple kinetic substeps, and fractional current contributions are accumulated at several positions along its guiding-center trajectory. The sum of these fractional contributions equals to the full marker weight, so that the total RE current is conserved. This procedure approximates a time average of the current deposition along the particle orbit and provides a smoother spatial representation than an instantaneous deposition at a single marker position~\cite{lopez_2025}
. In the present implementation, the parameter labeled "orbits" denotes the number of kinetic substeps used for orbit averaging. As shown in Figure{~}\ref{fig: total_comparison}, when only one orbit is used for orbit averaging, which corresponds to an effective convection speed of {$1\times10^{-3}c$}, the calculated RE current exhibits an abrupt jump during the late CQ phase. This jump is not interpreted as a physical instability or a sudden enhancement of RE generation. Instead, it is a numerical artifact caused by insufficient orbit averaging: the RE markers do not sample the orbit surface adequately before the background electromagnetic fields and source terms evolve appreciably, leading to a localized and non-smooth deposition of the RE current. When  orbit averaging is applied to at least five orbits, corresponding to $5\times10^{-3}c$, the artificial jump disappears and the total-current evolution becomes insensitive to further increases of the convection speed. Therefore, all drift-effect simulations discussed below use this converged orbit-averaging setting. 

\begin{figure*}[!htbp]
    \centering
    \subfloat[Marker orbits at $t=10\,\mathrm{ms}$\label{fig:orbit_10ms}]
    {\includegraphics[width=0.49\textwidth,trim=15 15 15 15,clip]{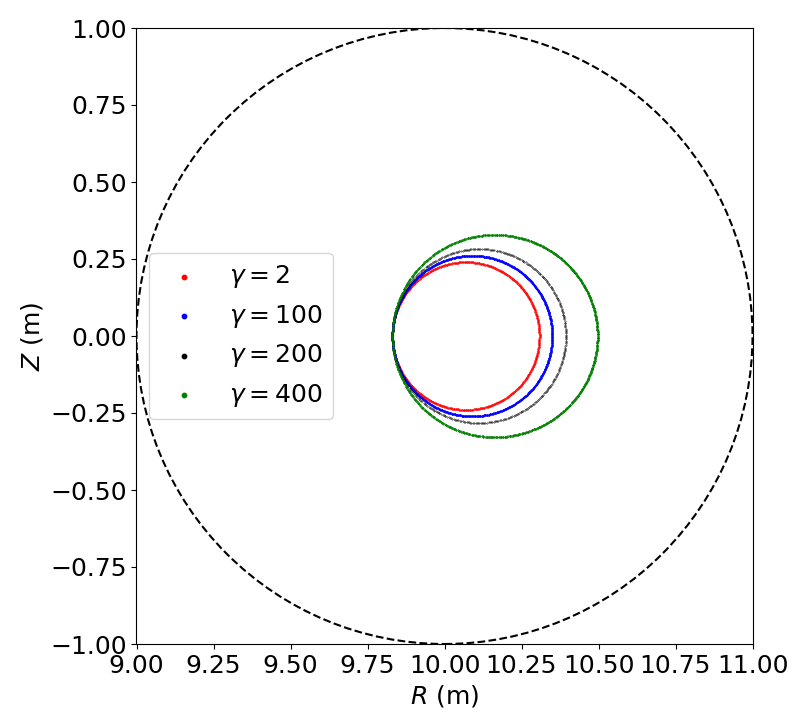}}
    \hfill
    \subfloat[Marker orbits at $t=45\,\mathrm{ms}$\label{fig:orbit_45ms}]
    {\includegraphics[width=0.49\textwidth,trim=15 15 15 15,clip]{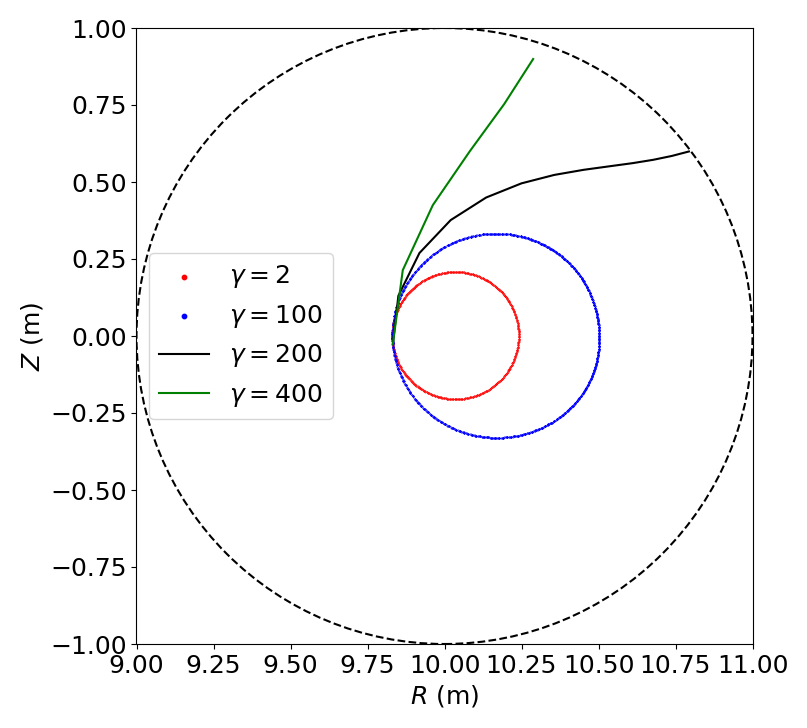}}

    \vspace{1em}

    \subfloat[Safety factor at $t=10\,\mathrm{ms}$\label{fig:q_10ms}]
    {\includegraphics[width=0.49\textwidth,trim=15 15 15 15,clip]{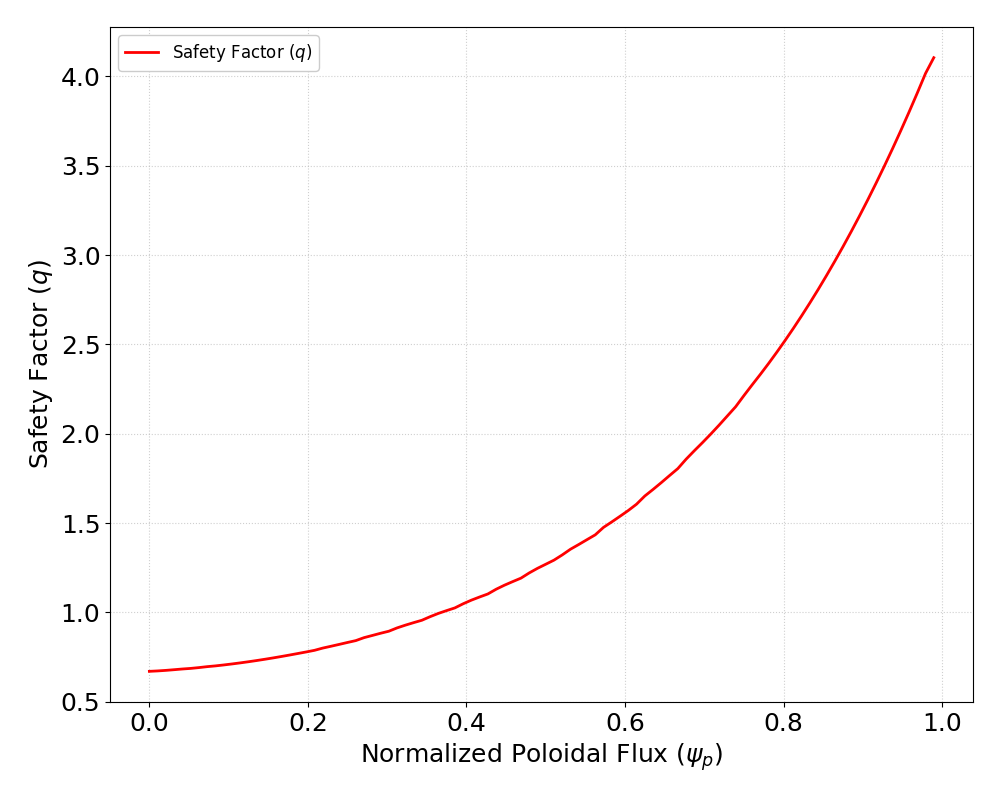}}
    \hfill
    \subfloat[Safety factor at $t=45\,\mathrm{ms}$\label{fig:q_45ms}]
    {\includegraphics[width=0.49\textwidth,trim=15 15 15 15,clip]{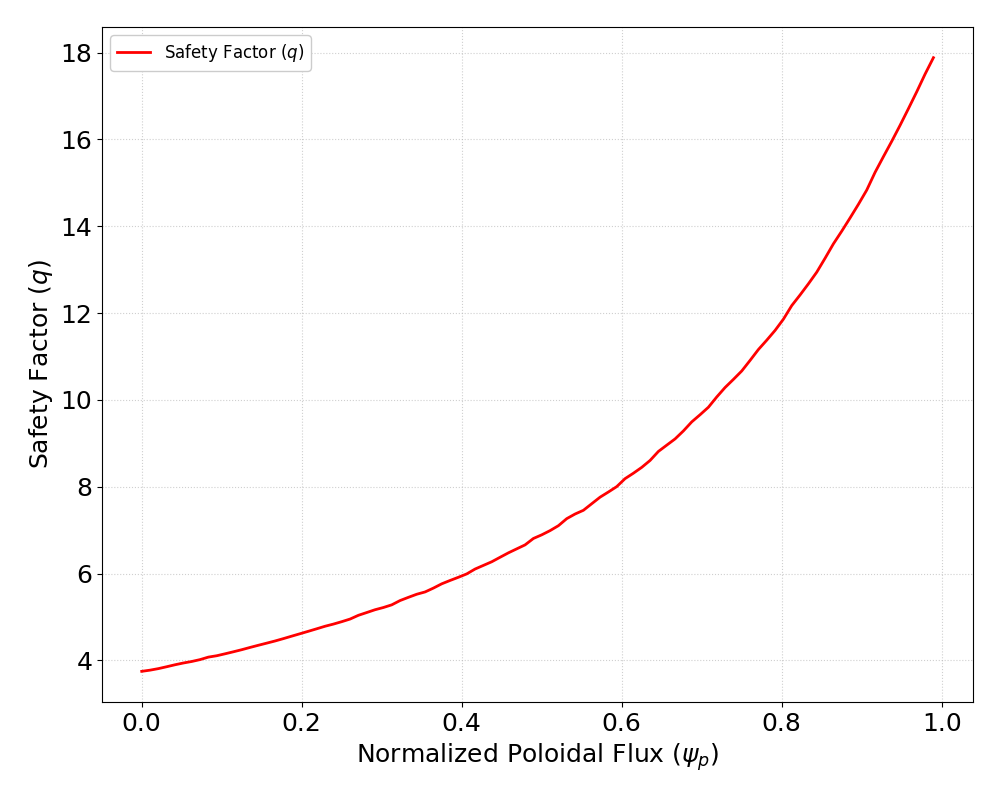}}

    \caption{(a)-(b) Test particle orbits with various energies and (c)-(d) the corresponding safety factor profiles at different times.}
    \label{fig:RE_orbits_and_safety_factor}
\end{figure*}

\begin{figure*}[!htbp]
    \centering{
        \includegraphics[width=1.2\textwidth]{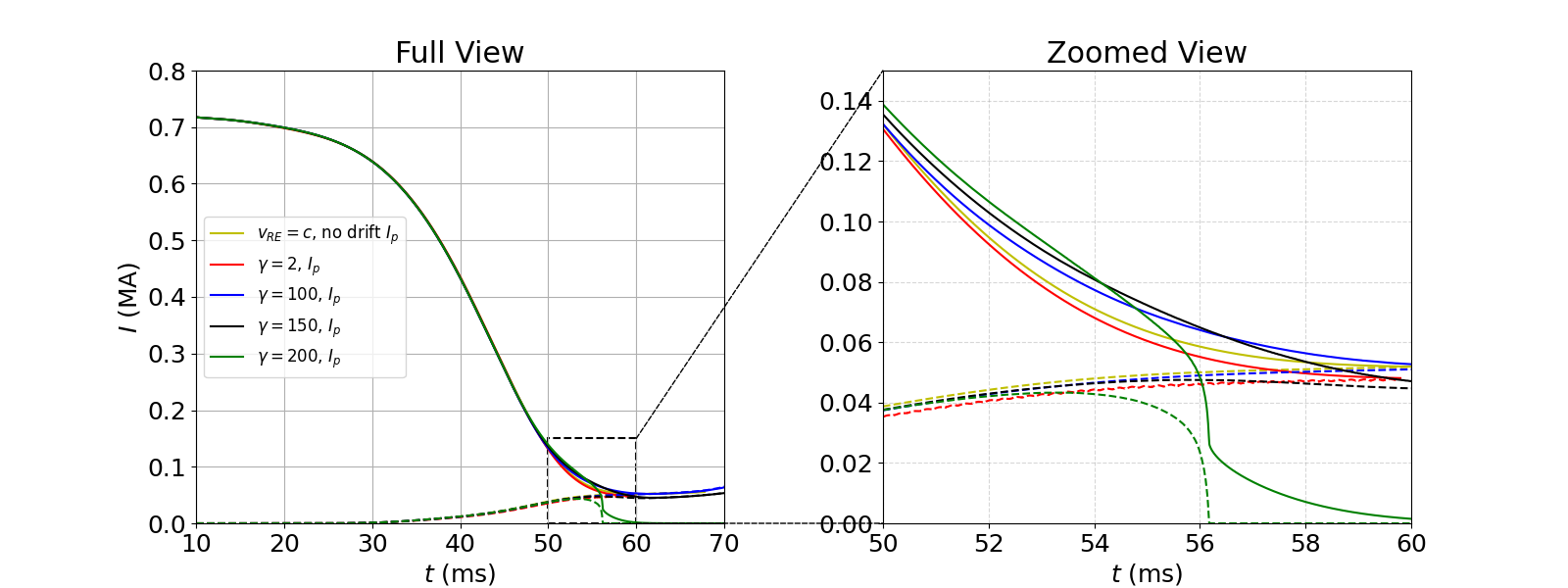}
    }
    \caption{Time evolution of total plasma and RE currents with various initial RE energies. The solid and dotted lines represent the total plasma current and total runaway current, respectively.}
        \label{fig: total current drift}
\end{figure*}

We ran the artificial quench cases with
$\gamma=2, 100, 150,$ and $200$, corresponding to kinetic energies of
approximately $0.5~\mathrm{MeV}$, $51~\mathrm{MeV}$,
$76~\mathrm{MeV}$, and $102~\mathrm{MeV}$, respectively. Figure \ref{fig:orbit_10ms} and Figure \ref{fig:orbit_45ms} show that the larger the electron energy is, the wider the drift orbit becomes. Figure \ref{fig: total current drift} shows how the total toroidal plasma current evolution varies with the runaway electron energy, in good agreement with Ref.~\cite{abdullaev_2016}. For RE energies below about $100~\mathrm{MeV}$, the total current evolution
changes only moderately. However, when the RE energy becomes sufficiently high, as illustrated in the $\gamma=200$ case, the RE current is rapidly lost after reaching its maximum, and no sustained runaway-current plateau is found.

\begin{figure*}[!htbp]
    \centering
    \subfloat[$t=25.0\mathrm{ms}$]{
        \includegraphics[width=0.4\textwidth]{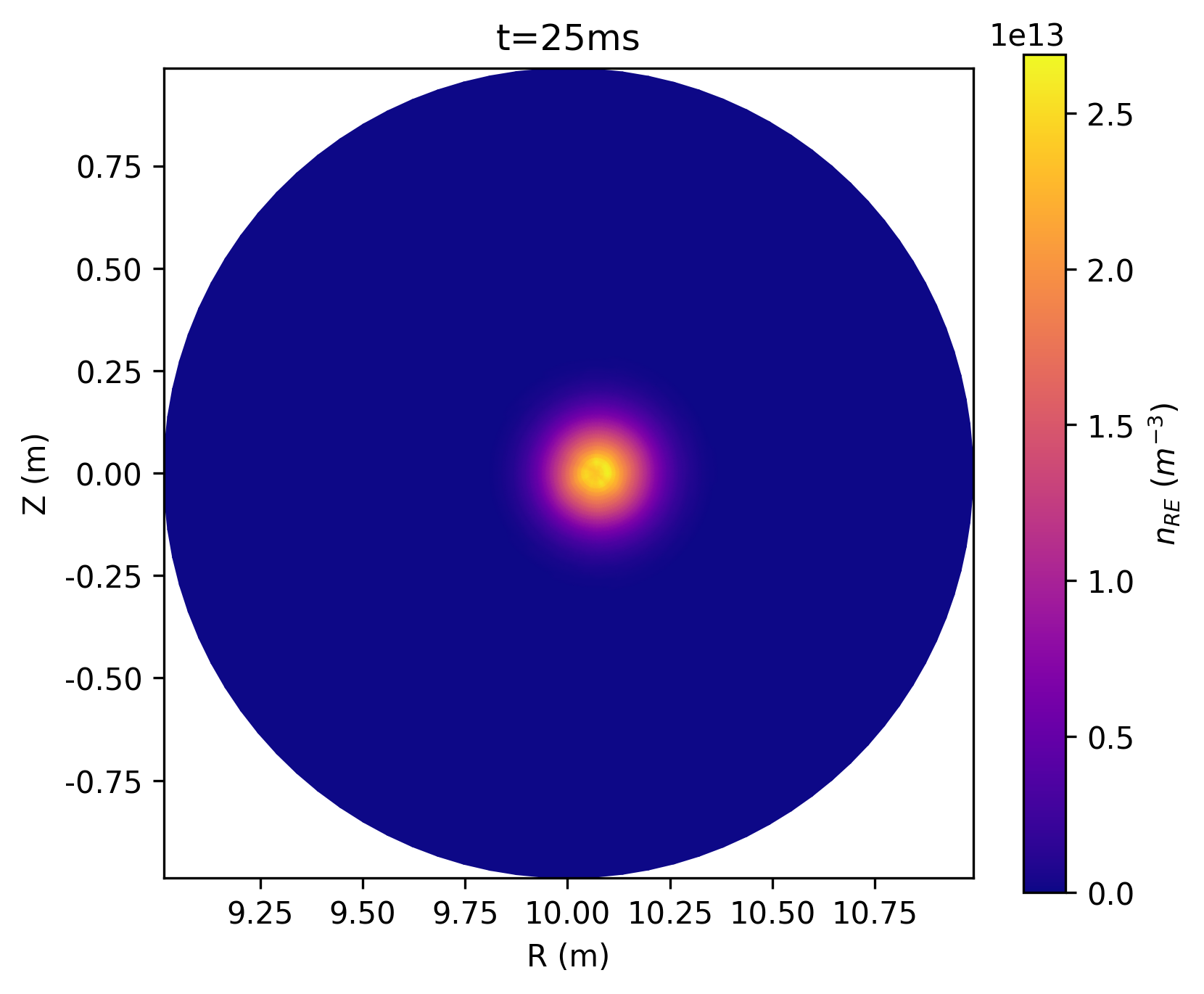}
        \label{fig: RE con 25.0ms}
    }
        \centering
    \subfloat[$t=49.7\mathrm{ms}$]{
        \includegraphics[width=0.4\textwidth]{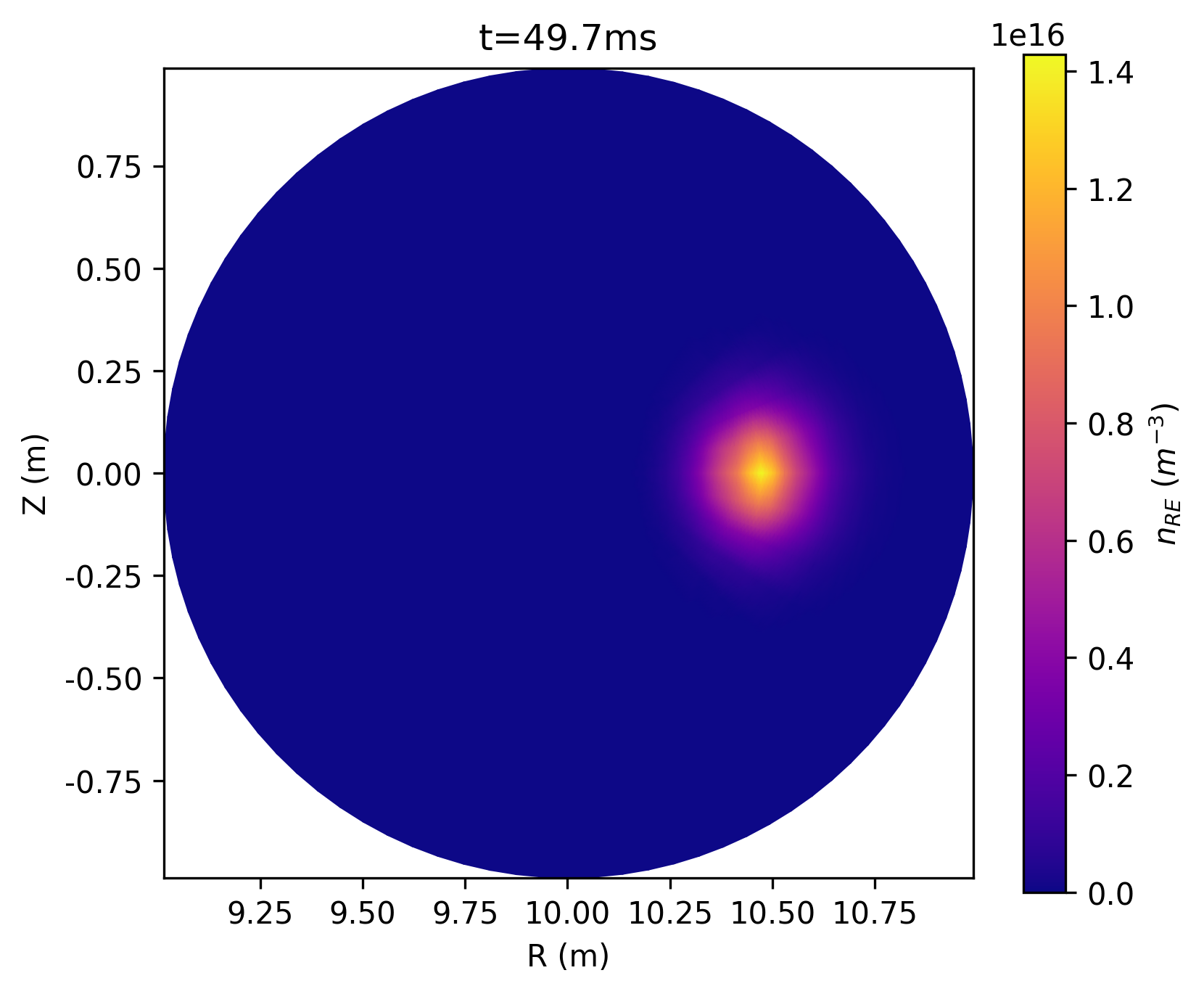}
        \label{fig: RE con 49.7ms}
    }
        \hfill
    \subfloat[$t=52.0\mathrm{ms}$]{
        \includegraphics[width=0.4\textwidth]{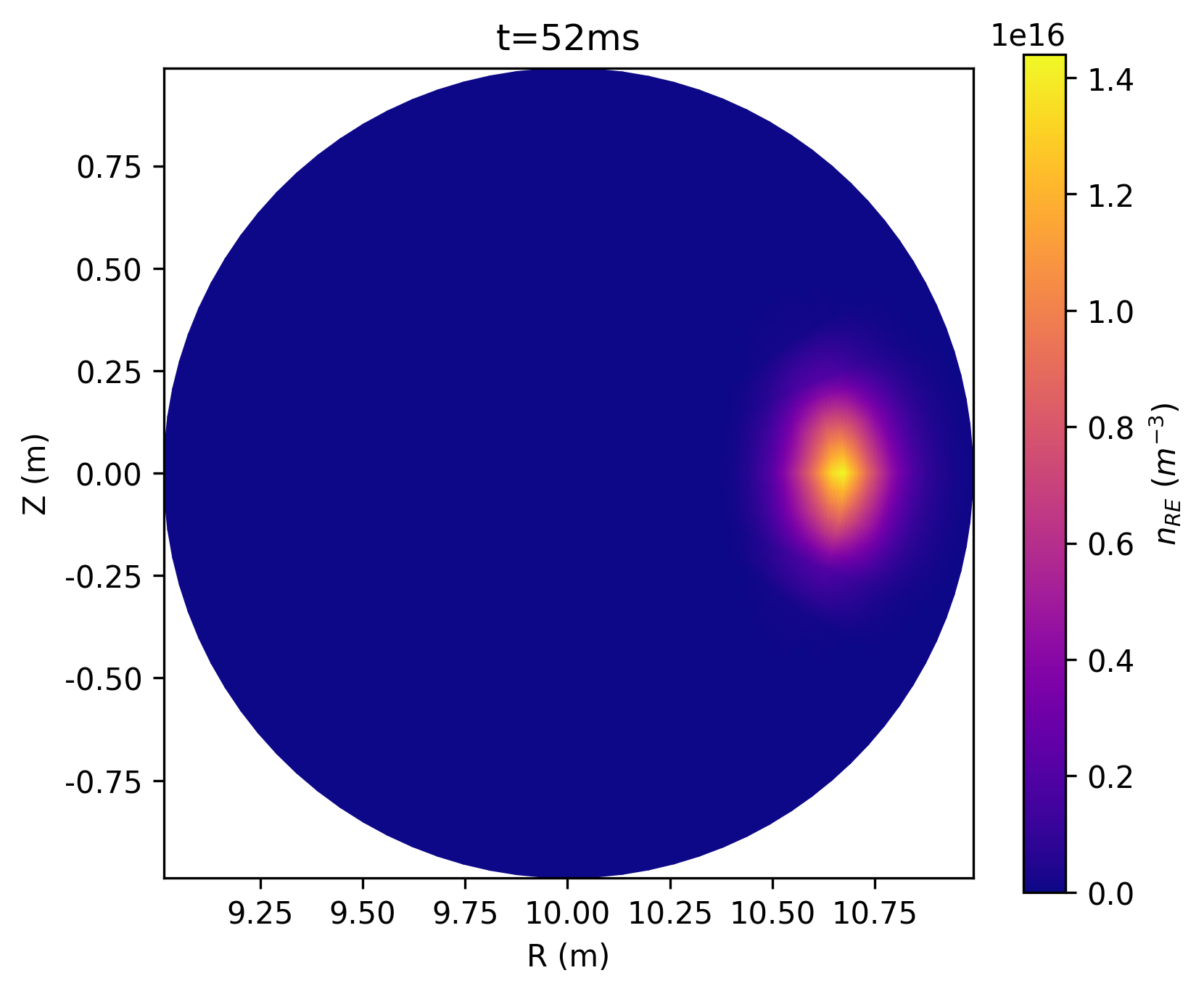}
        \label{fig: RE con 52.0ms}
    }
        \centering
    \subfloat[$t=55.7\mathrm{ms}$]{
        \includegraphics[width=0.4\textwidth]{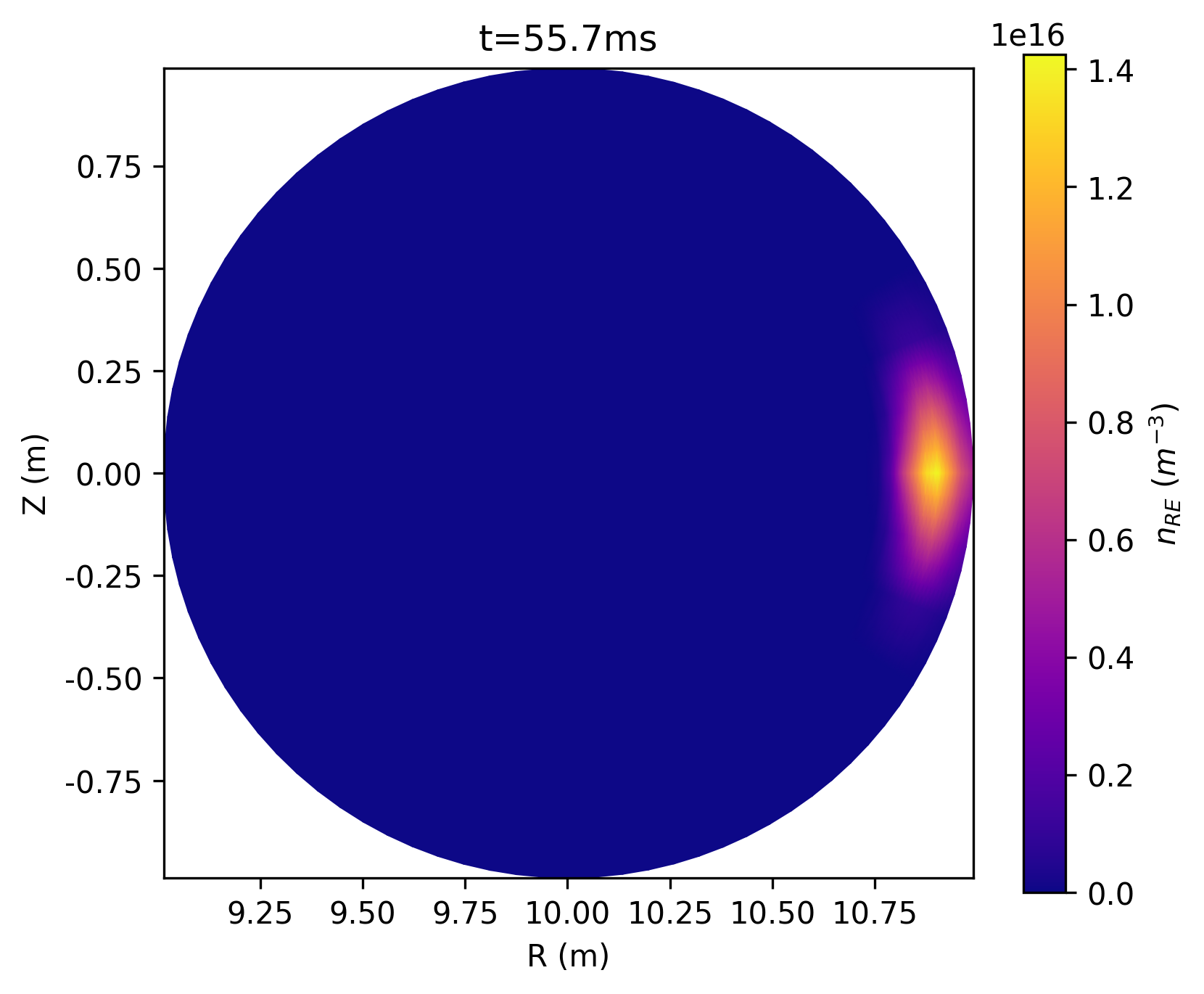}
        \label{fig: RE con 55.7ms}
    }
    \caption{RE density contours {in} the {poloidal plane} with $\gamma=200$ at (a) $t=25.0\,\mathrm{ms}$, (b) $49.7\,\mathrm{ms}$, (c) $52.0\,\mathrm{ms}$, and (d) $55.7\,\mathrm{ms}$. }
    \label{fig: nRE contour gamma=200}
\end{figure*}

\begin{figure*}[!htbp]
    \centering
    \subfloat[RE density 1D profile at t=40ms]{
        \includegraphics[width=0.8\textwidth]{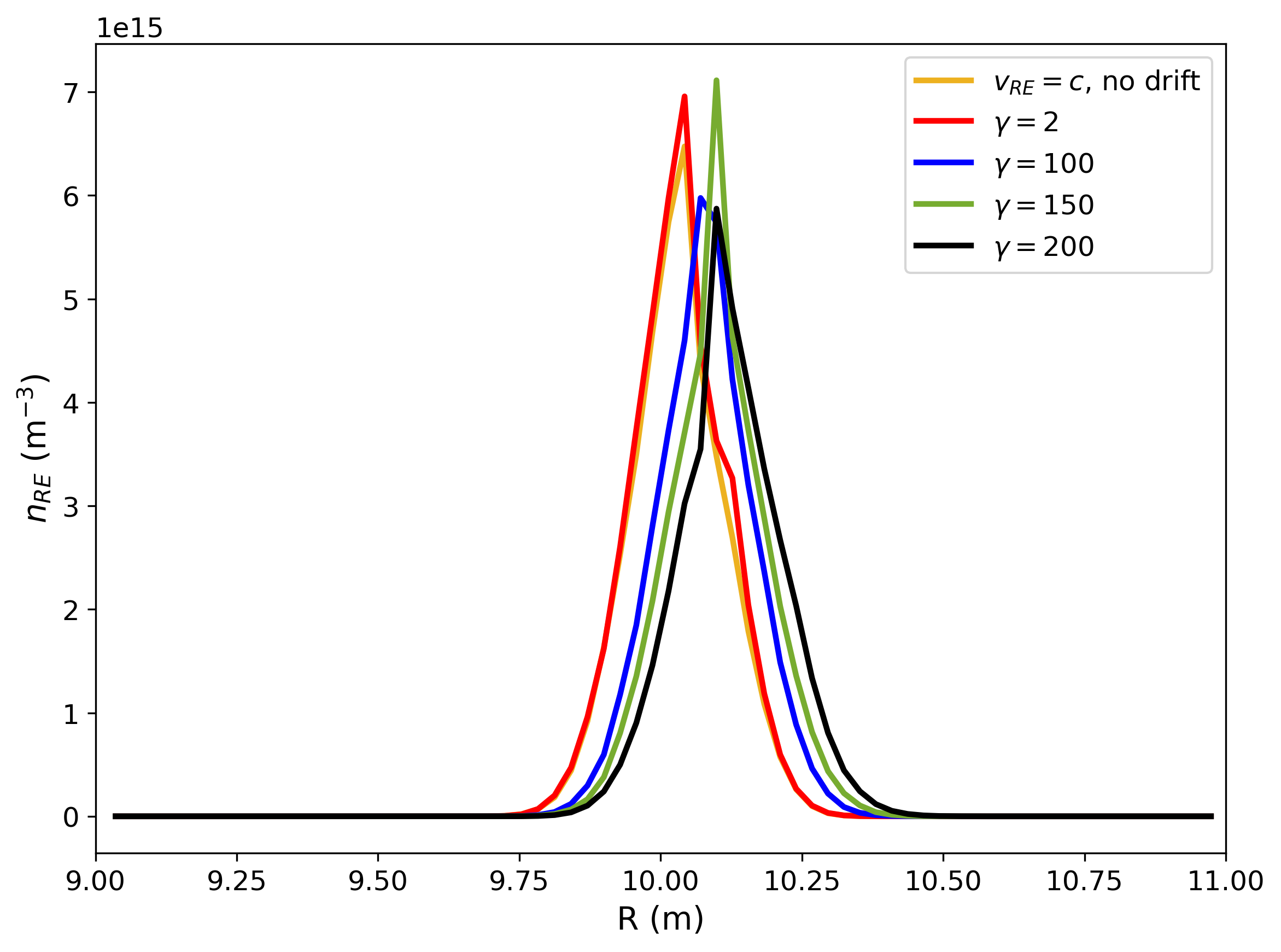}
        \label{fig: RE density 1D profile at t=40ms}
    }
    \hfill
    \subfloat[Final RE density 1D profile]{
        \includegraphics[width=0.8\textwidth]{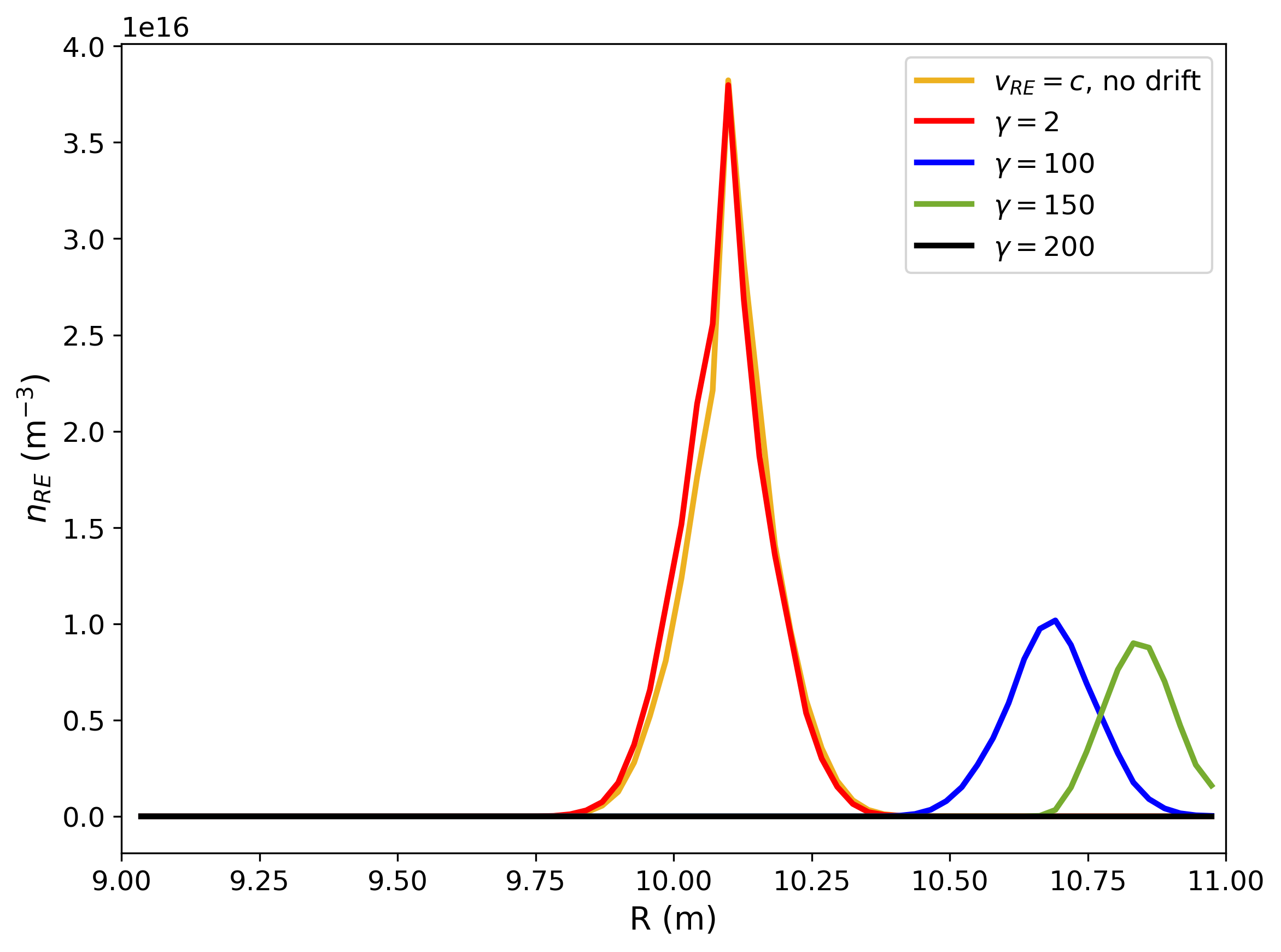}
        \label{fig: Final RE density 1D profile}
    }
    \caption{RE density profiles as functions of major radius for various energies at (a) $t=40\,\mathrm{ms}$ and (b) during the post-CQ stage.}
    \label{fig:RE density 1D profile}
\end{figure*}

Figure \ref{fig: nRE contour gamma=200} shows that for $\gamma=200$, REs move toward the plasma edge as the total runaway current decreases. This transport mechanism is not due to open magnetic field lines, but rather due to the widening of drift orbits, as shown in Figure \ref{fig:RE_orbits_and_safety_factor}. At the beginning of the simulation, the safety factor is small, so the REs are well confined. As the current quench (CQ) starts, the safety factor increases rapidly, and the drift surfaces expand. Consequently, the high-energy runaway electrons are eventually lost to the edge. The plasma current continues to decrease, leading to the complete loss of all REs. Although the total currents for the $\gamma=2$ and 100 cases do not show big differences, Figure \ref{fig:RE density 1D profile} indicates that the RE energy has a remarkable influence on the RE distribution.

\section{Discussion and conclusions}
\label{sec:dis}
 In summary, we have developed a PIC model for runaway electron generation and dynamics, which is coupled with the 3D fully nonlinear  MHD code NIMROD. We have implemented two methods of RE generation: one is directly adding markers, and the other is changing the weight of existing markers, which are proven to be equivalent  and consistent. We have also included the guiding-center drifts to the RE parallel motion which would enable future studies on RE transport in stochastic magnetic fields.

 The simulations of RE current plateau formation during an artificial CQ have been conducted using the NIMROD code including the PIC-RE model which has been benchmarked with fluid-RE models in DREAM, JOREK, and the NIMROD codes. The time evolution of the total current and the post-CQ current profiles show good agreement among the models. The inclusion of RE drifts leads to significant changes in the RE distribution and plateau evolution. The lower-energy REs, such as in the $\gamma=2$ and $\gamma=100$ cases, remain
relatively well confined during the CQ, whereas sufficiently energetic REs,
especially in the $\gamma=200$ case, exhibit significantly widened drift
orbits and can eventually be lost due to the increase of the $q$ value.

 If we consider realistic disruption processes, the effects of stochastic fields and the finite aspect ratios must be taken into account. The influences of drifts could be even more significant in these situations. Future studies plan to use our PIC-RE model in NIMROD code to study more realistic CQ scenarios and  RE distributions, while kinetic effects such as the acceleration in the electric field, collisions, and radiation will be considered.

\begin{acknowledgments}
 The authors are grateful for the supports from
the NIMROD team and the J-TEXT team. This work was supported by the National MCF Energy R\&D Program of China (Grant
No. 2019YFE03050004), the  U.S. Department of Energy (Grant No. DEFG02-86ER53218) and the Hubei International Science and Technology Cooperation
Project under Grant No. 2022EHB003. This research used
resources of the National Energy Research Scientific Computing Center, a DOE Office of
Science User Facility supported by the Office of Science of the U.S. Department of Energy
under Contract No. DE-AC02-05CH11231 using NERSC award FES-ERCAP0027638. The computing work in this paper was
also supported by the Public Service Platform of High Performance Computing by Network
and Computing Center of HUST.  
 \end{acknowledgments}

\section*{AUTHOR DECLARATIONS}

The authors have no conflicts to disclose.

\section*{DATA AVAILABILITY}

The data that support the findings of this study are available from the corresponding author upon reasonable request.

\bibliography{ref}

\clearpage

\clearpage

\clearpage

\end{document}